\documentclass{article}

\usepackage{arxiv}

\usepackage[utf8]{inputenc} % allow utf-8 input
\usepackage[T1]{fontenc}    % use 8-bit T1 fonts
\usepackage{hyperref}       % hyperlinks
\usepackage{url}            % simple URL typesetting
\usepackage{booktabs}       % professional-quality tables
\usepackage{amsfonts}       % blackboard math symbols
\usepackage{nicefrac}       % compact symbols for 1/2, etc.
\usepackage{microtype}      % microtypography
\usepackage{lipsum}
\usepackage{graphicx}
\usepackage{subcaption}
\graphicspath{ {./images/} }
\usepackage{amsmath}

\title{Research on the Acoustic Emission Source Localization Methodology in Composite Materials based on Artificial Intelligence}

\author{
 Jongick Won \\
  Dept. of Automotive Engineering\\
  Hanyang University\\
  \texttt{jongickwon@gmail.com} \\
   \And
 Hyuntaik Oh \\
  Dept. of Automotive Engineering\\
  Hanyang University\\
  \texttt{oht0709@hanyang.ac.kr} \\
  \And
 Jae Sakong \\
  Hanwha Aerospace\\
  \texttt{dooboo@hanyang.ac.kr}
}

\begin{document}
\maketitle
\begin{abstract}
In this study, methodology of acoustic emission source localization in composite materials based on artificial intelligence was presented. Carbon fiber reinforced plastic was selected for specimen, and acoustic emission signal were measured using piezoelectric devices. The measured signal was wavelet-transformed to obtain scalograms, which were used as training data for the artificial intelligence model. AESLNet(acoustic emission source localization network), proposed in this study, was constructed convolutional layers in parallel due to anisotropy of the composited materials. It is regression model to detect the coordinates of acoustic emission source location. Hyper-parameter of network has been optimized by Bayesian optimization. It has been confirmed that network can detect location of acoustic emission source with an average error of 3.02mm and a resolution of 20mm.
\end{abstract}

% keywords can be removed
\keywords{Structural Health Management \and Artificial Intelligence \and Acoustic Emission \and Composite Material}

\section{Introduction}
~Composites, which are formed by the integration of reinforcement and matrix, allow for flexible material design optimization for its usage by adjusting the combination of the reinforcement and matrix. Composites are continuously being explored for application in diverse fields, especially in areas where need lightweight constructions, notably in aerospace and automotive sectors. Fiber reinforced plastics(FRP) are widely utilized, and they inherently display anisotropy because of their structural features. Also, because matrix and fiber have different mechanical properties, they fail under different conditions. This makes complicated failure behavior, and challenging to accurately determine the onset of failure. These damages degrade the mechanical properties such as strength and stiffness of the material. Structural health management is essential to ensure the reliability of structure.
The acoustic emission(AE) test is extensively used to detect and monitor damage in composites that are hard to recognize in visually[1-3]. The AE test is a method that measures the energy released when the material is damaged or deformed in the elastic wave form. Consequently, it is possible to localize the damage by identifying the location of the AE source. Previous research on localizing the AE source has primarily focused on measuring the time of arrival on multiple sensors[4-6]. This approach is generally assuming a propagation speed is constant. However as previously mentioned, composites consist of a combination of reinforcement and matrix making it difficult to calculate the exact propagation path and speed[7]. These structural characteristic are the factor that reduce the accuracy of this approach. Nowadays, various features other than arrival time are used to localize the AE  sources. Approaches based on features form the AE signal can be divided into features from time domain[8,9], and features from frequency domain[10,11]. Time-domain feature-based approaches are possible to reflect the time continuity of the signal. But it is difficult to understand the characteristics of signal intuitively. Conversely, frequency-domain feature-based approaches, while the aspect of signal is more understandable, it is challenging to incorporate temporal continuity. To analyze non-stationary signals such as AEs, methods capable of reflecting temporal continuity and transforming signals into the frequency domain, such as short-time Fourier transform(STFT) and Wavelet transform, are used[10,11]. As mentioned earlier, composites are not homogeneous medium, wave propagation speeds are non-linear in frequency domain[7]. To accurately localize the AE source in composite, it is necessary to consider the response characteristics in the frequency domain. Recently, neural networks have been widely employed for non-linear causal inference and are also being attempted in the field of structural health management[8-11]. However, most of research on damage detection based on artificial intelligence applies the same convolution kernel to signals measured from multiple sensors. Because of the anisotropic nature of composites, this method is hard to reflect the sensor response difference accurately based on their positions. We propose the AESLNet(Acoustic Emission Source Localization Network) which can overcome the weakness of previous research.
Firstly, AESLNet use scalogram as input data to consider the temporal continuity in frequency domain. Secondly, by passing the signals measured from each sensor through different convolutional layers, the response characteristics due to the sensor positions were trained during training. Finally, optimize the hyper parameter to maximize the performance of trained model.

 \begin{figure} [!b]
    \centerline{\includegraphics[width=0.7\linewidth]{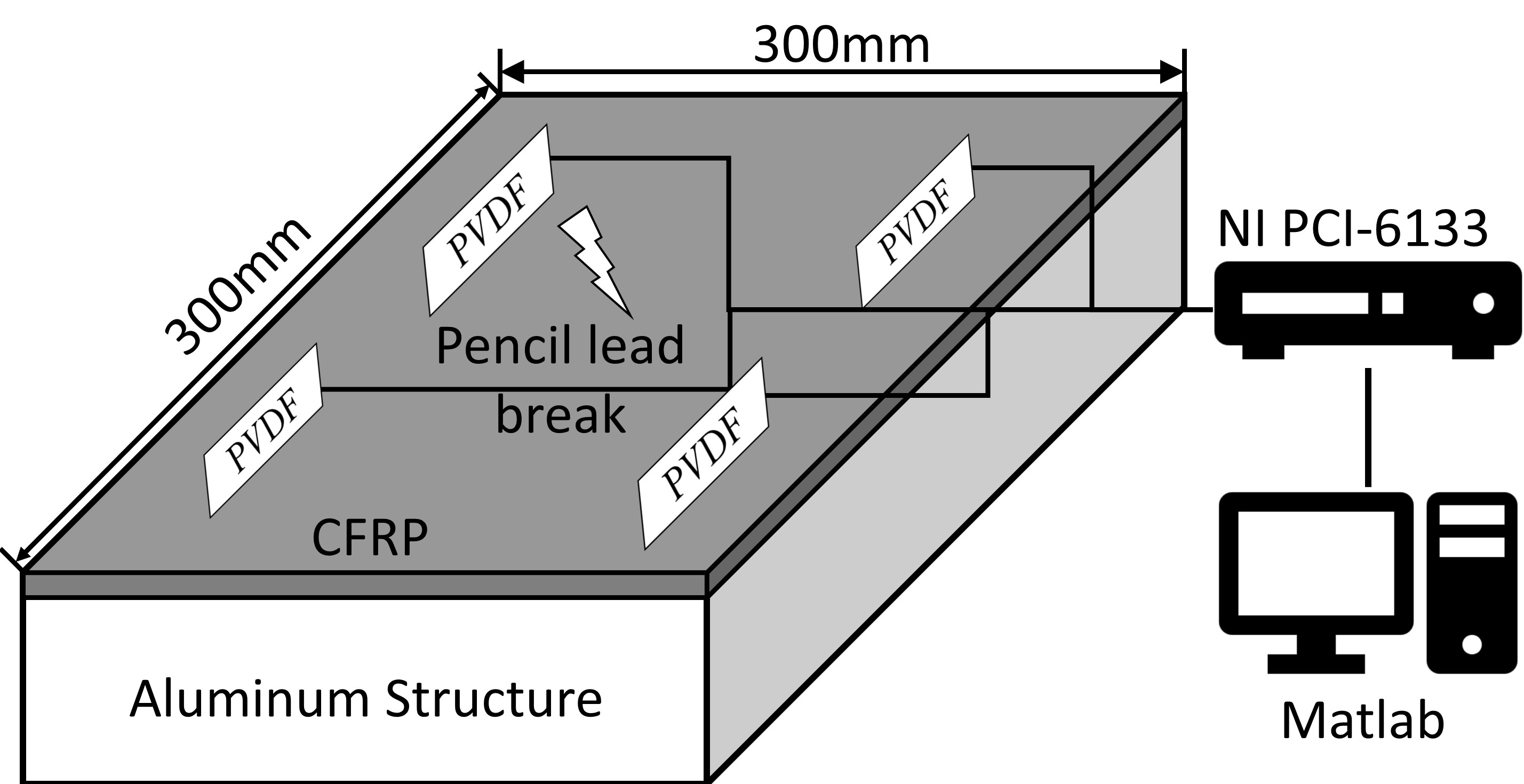}}
    \caption{Schematic of experiment}
    \label{figure_1}
 \end{figure}

\begin{figure}[!b]
    \centering
    \begin{subfigure}[b]{0.4\textwidth} % Adjust width as needed
        \centering
        \includegraphics[height=\textwidth]{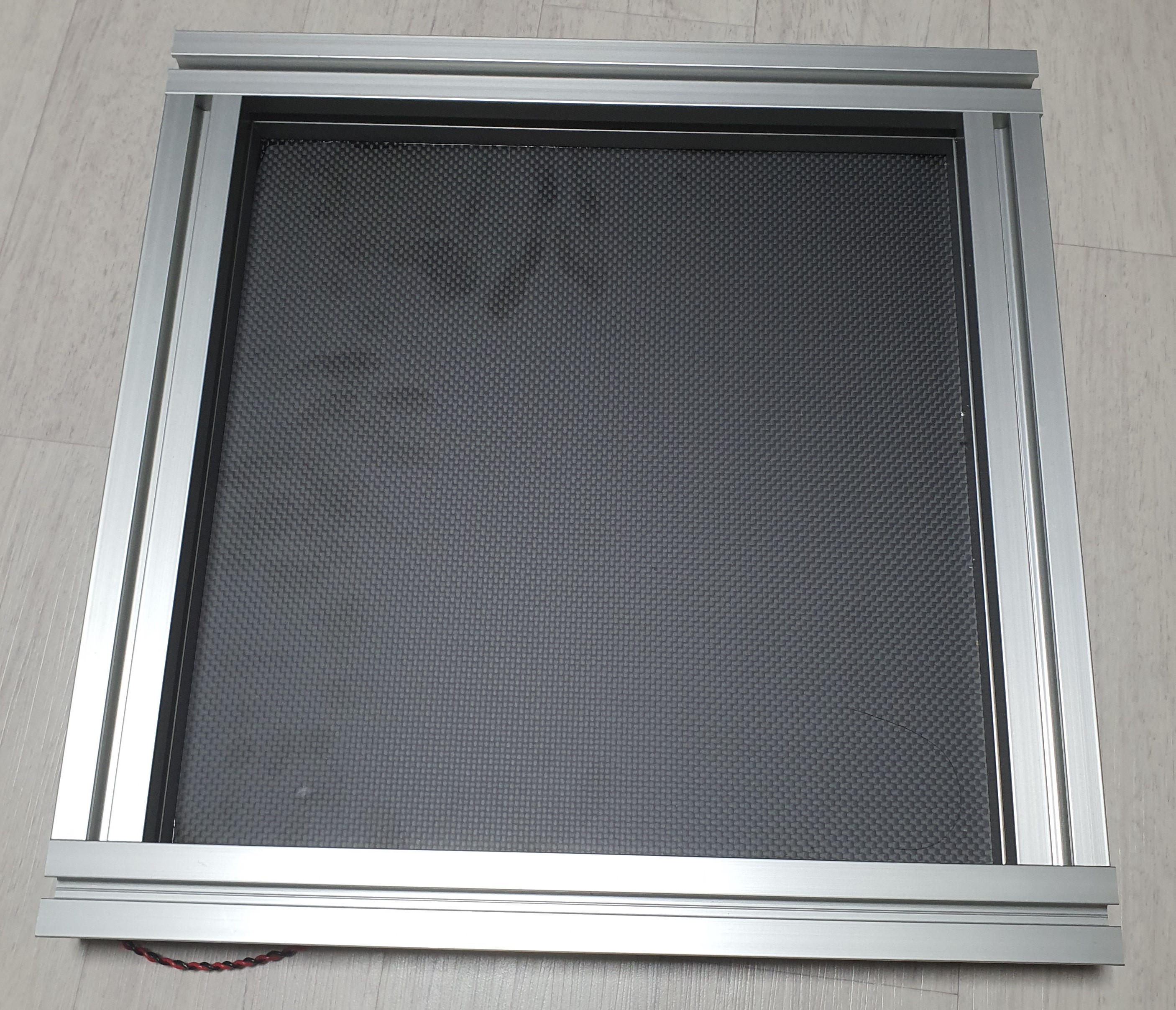}
        \caption{The back of the specimen}
        \label{fig:subfig1}
    \end{subfigure}
    \hspace{15 mm}
    \begin{subfigure}[b]{0.4\textwidth} % Adjust width as needed
        \centering
        \includegraphics[height=\textwidth]{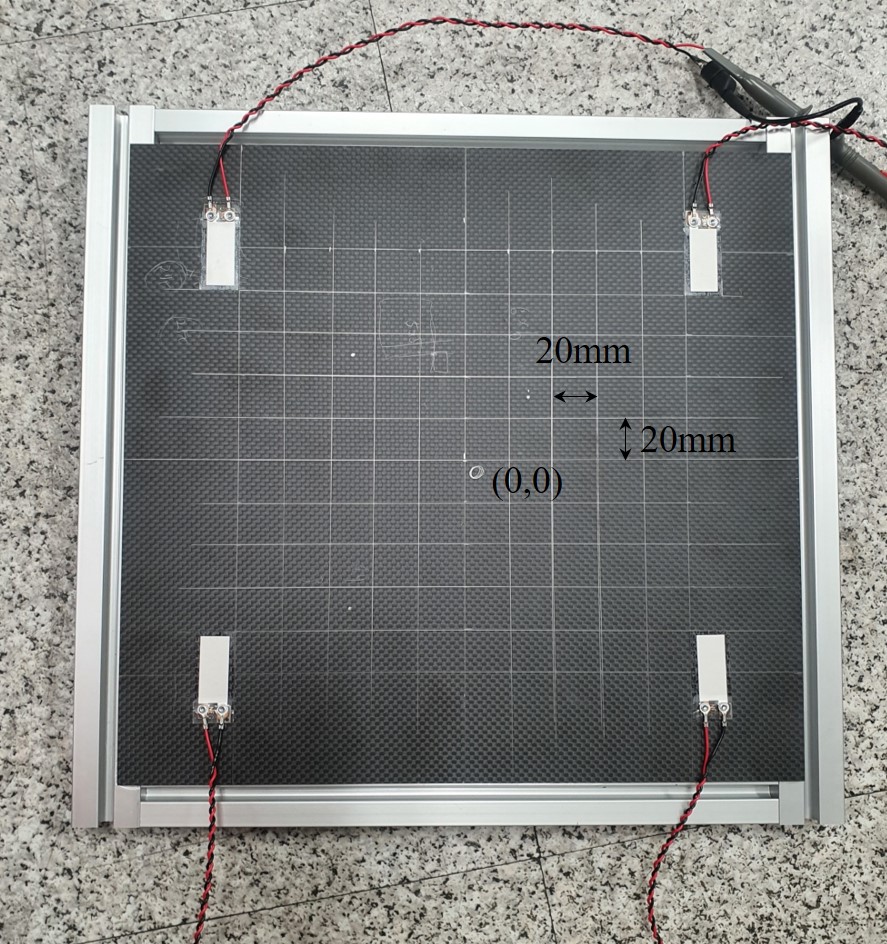}
        \caption{Coordinate setting of specimen}
        \label{fig:subfig2}
    \end{subfigure}
    \caption{Experimental Setup}
    \label{fig:experimental_setup}
\end{figure}

\section{Materials and methods}
\label{sec:headings}
\subsection{Experiments}
 ~Hsu-Nielson pencil lead breakage test(PLB) was performed to simulate the AE signal. Hsu devised a PLB to simulating the AE signal[12], which has been applied in various AE signal research[13,14].
 In this study, experiments has been conducted as figure 1 to get AE signal dataset. 300 $\times$ 300 $\times$ 3mm sized carbon fiber reinforced plastic(CFRP) is used as specimen, and aluminum structure is used to fix each edge of the specimen. Polyvinylidene fluoride(PvDF) has installed each corner of the specimen to measure the AE signal as figure 2(a). Measured AE signals are discretized by data acquisition(National Instrument PCI-6133) in 1Msps.
 ~In order to acquire AE signal data from various positions on the specimen, PLB test has conducted on each point of 9x7 grid set on the specimen as shown in figure 2(b). 189 train data, and 15 validation data has been gathered. Every test were performed at a 45-degree and 2mm length of the lead.

\subsection{Data processing}
%\subsubsection{Signal Processing}
 ~Signal processing were performed to extract the feature of the AE signal, as shown in figure 3. The wavelet transform is one of the method to analyze the non-stationary signal in time-frequency domain. Wavelet transform is a method to decompose the signal into multiple mother wavelets[15-18]. Mother wavelet is a finite-length waveform that represents translated and scaled copies. Wavelets are effective for the decomposition and analysis of non-stationary signals due to their ability to provide both high resolution in time and frequency. This makes wavelet transform appropriate for decomposition of AE signals which are characterized by high frequencies and non-stationary behavior. The wavelet transform can be formulated as shown below.

\begin{figure}[t]
    \centering
    \includegraphics[width=0.5\linewidth]{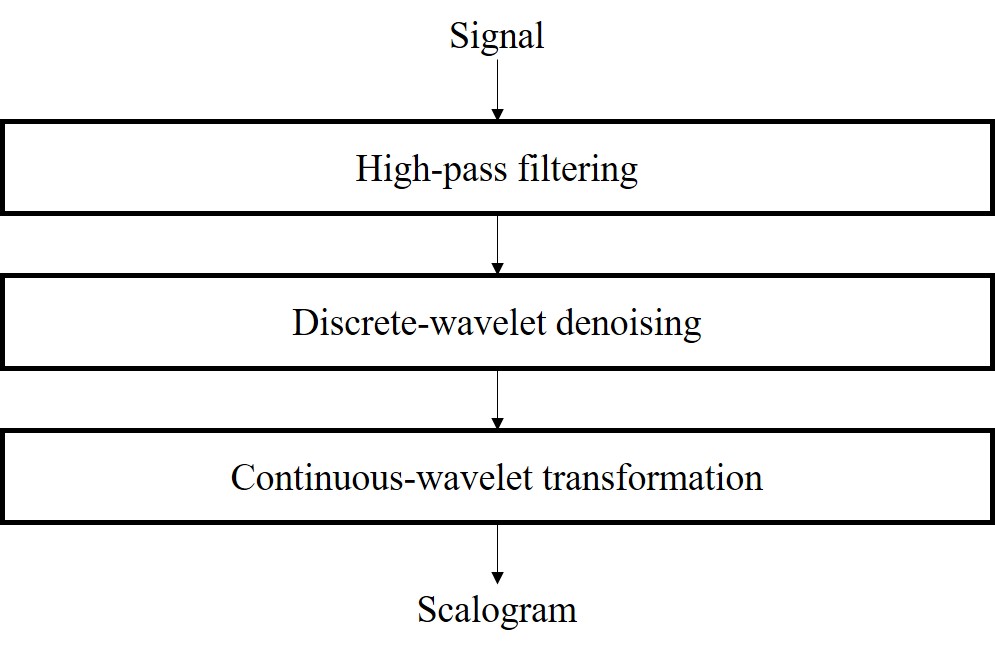}
    \caption{Signal Processing}
    \label{fig:signal_processing}
\end{figure}

\begin{figure}[t]
    \centering
    \begin{subfigure}[b]{0.45\linewidth} % adjust width as needed
        \centering
        \includegraphics[height=0.8\linewidth]{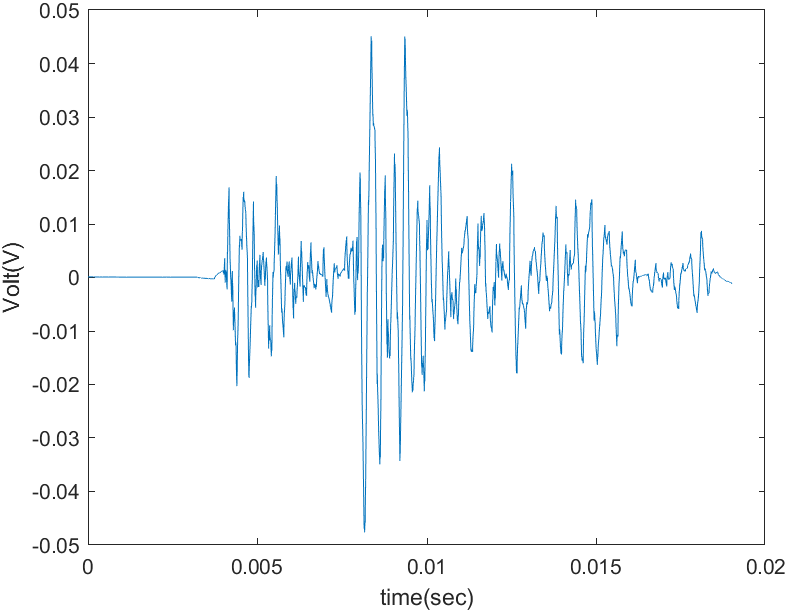}
        \caption{Signal measured by PVDF}
        \label{fig:pvdf_signal}
    \end{subfigure}
    \hfill
    \begin{subfigure}[b]{0.45\linewidth} % adjust width as needed
        \centering
        \includegraphics[height=0.8\linewidth]{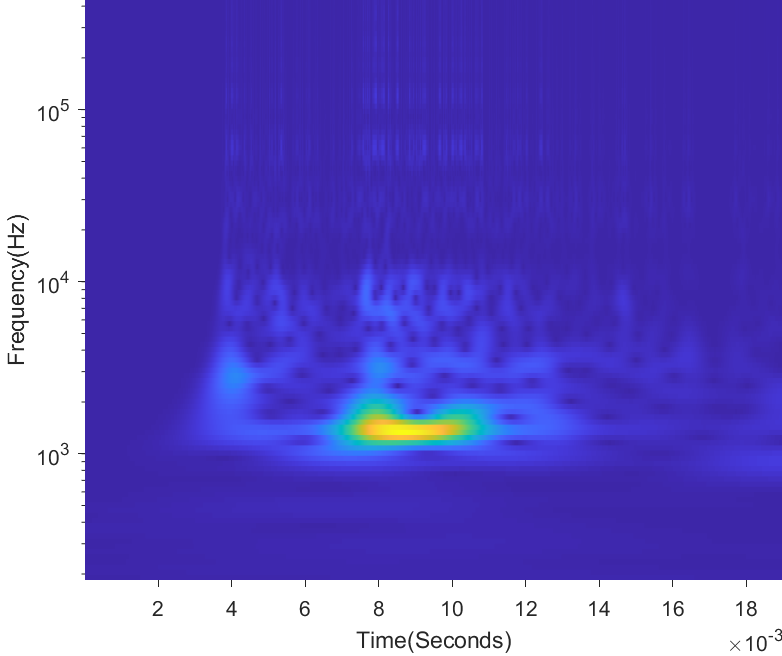}
        \caption{Scalogram of acoustic emission signal}
        \label{fig:scalogram}
    \end{subfigure}
    \caption{Acoustic emission signal}
    \label{fig:acoustic_emission}
\end{figure}

\begin{equation}
[W_{\psi}f](a,b)= { {1} \over {\sqrt{ |a|}}} \int_{-\infty}^{\infty} \Bar{\psi \left(  {{x-b} \over {a}}   \right) f(x)dx}
\end{equation}
~Where $a$ is scale factor and $b$ is shifting factor.
 Process of preprocessing is shown in figure 3 to get data for input of AESLNet. Measured signal passed 1kHz high pass filter and de-noised by discrete wavelet transform in order to isolate the AE signals as figure 4(a). The AE signals were converted into scalograms as figure 4(b), which represents behavior of the signals over time within frequency domain.

%\subsubsection{Data Processing}
 ~Four scalograms of AE signals are measured from 4 sensors located each corner of the specimen from one PLB test. Four scalogram is converted to 4 channel image data as figure 5 to use input of the AESLNet. Data divided into 189 train data and 15 test data.
 
\begin{figure}
    \centering
    \includegraphics[width=0.7\linewidth]{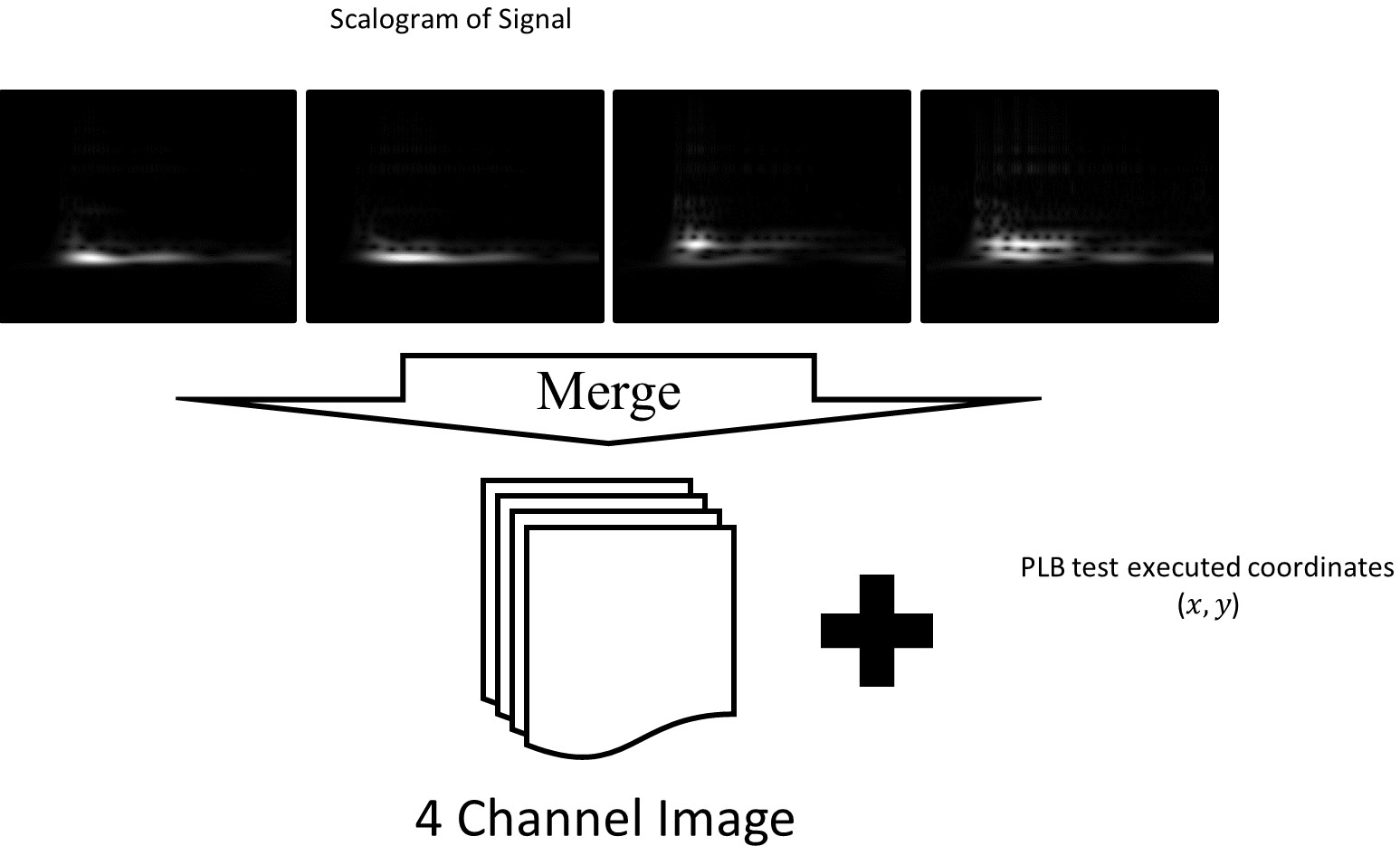}
    \caption{Method to construct the database}
    \label{fig:enter-label}
\end{figure}

\subsection{Acoustic Emission Signal Propagation Speed Calculation}
 To compare the accuracy of AESLNet with method based on wave propagation speed, wave propagation speed has been calculated. Propagation speed is calculated by follow equations.
 \begin{equation}
v_{ij} = {{l_i - l_j} \over {t_i-t_j}},~~~i,j=1,2,3,4
 \end{equation}
Where $l_i$ is length from signal source to $i$th sensor, $t_i$ is propagation time. Across 189 experiments, propagation speeds were assessed for differing combinations of sensor placements, cumulating in 1,134 unique instances of propagation speed calculations.

\section{Acoustic Emission Source Localization Network}
\subsection{Architecture of AESLNet}
~ The architecture of AESLNet is shown in figure 6. AESLNet consist of four feature extraction module and detection module. This aims to reflect the non-linear propagation characteristics of AE signals according to spatial positions into the training phase. Each channel was fed into a different convolution module to extract the features of each signals in feature extraction module. Extracted features combined in detection module to predict the coordinates of the source of AE signals.
~The feature extraction modules are consisting of four convolution layers, four max pooling layers, one dropout layer, and one fully connected layer. The detection module includes one batch normalization layer and five fully connected layer. The module outputs the continuous coordinates (x, y) representing the position of the signal source. 
\begin{figure}[htbp]
    \centering
    \includegraphics[width=1\linewidth]{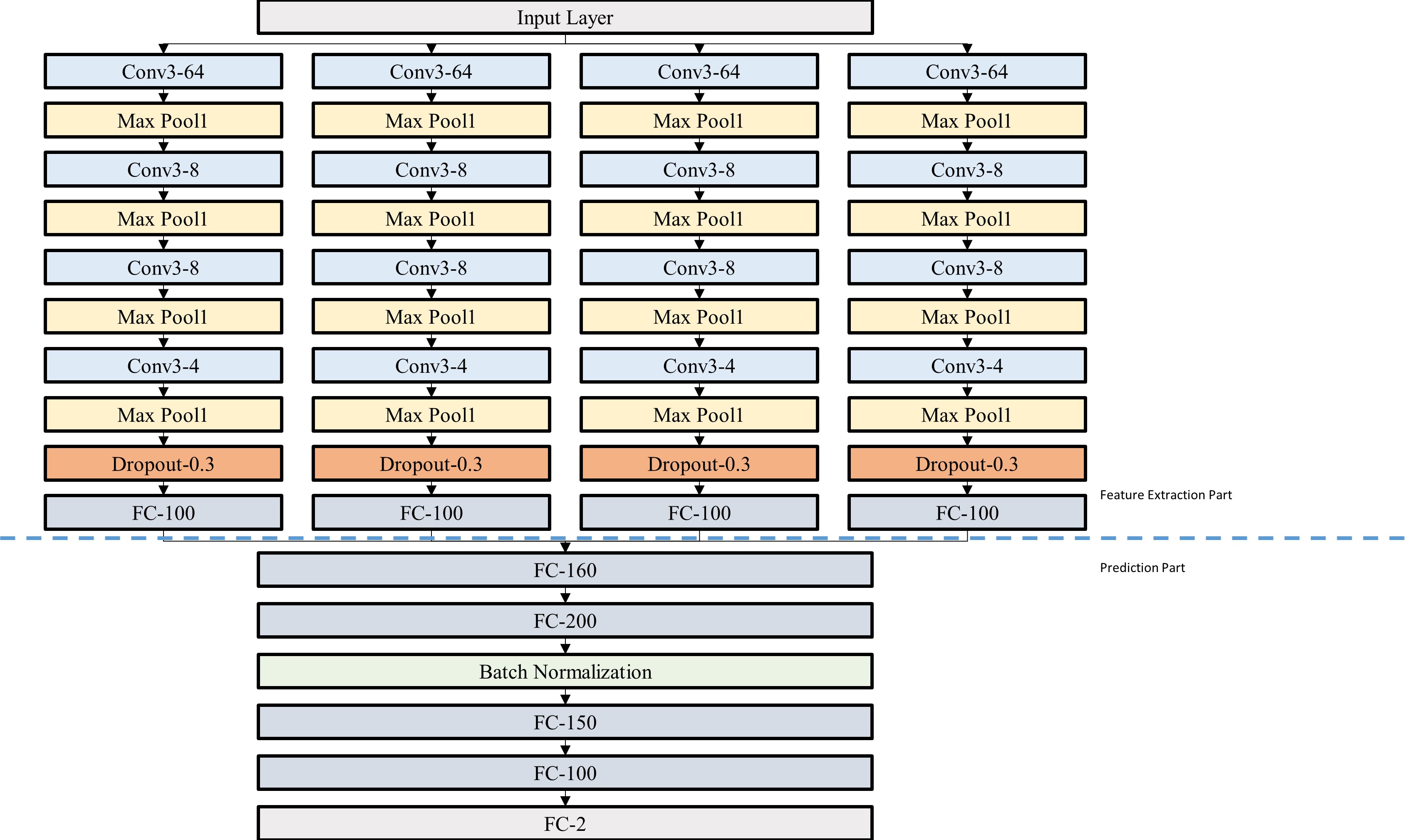}
    \caption{Overall structure of AESLNet}
    \label{fig:enter-label}
\end{figure}

\begin{figure}[htbp]
    \centering
    \includegraphics[width=0.7\linewidth]{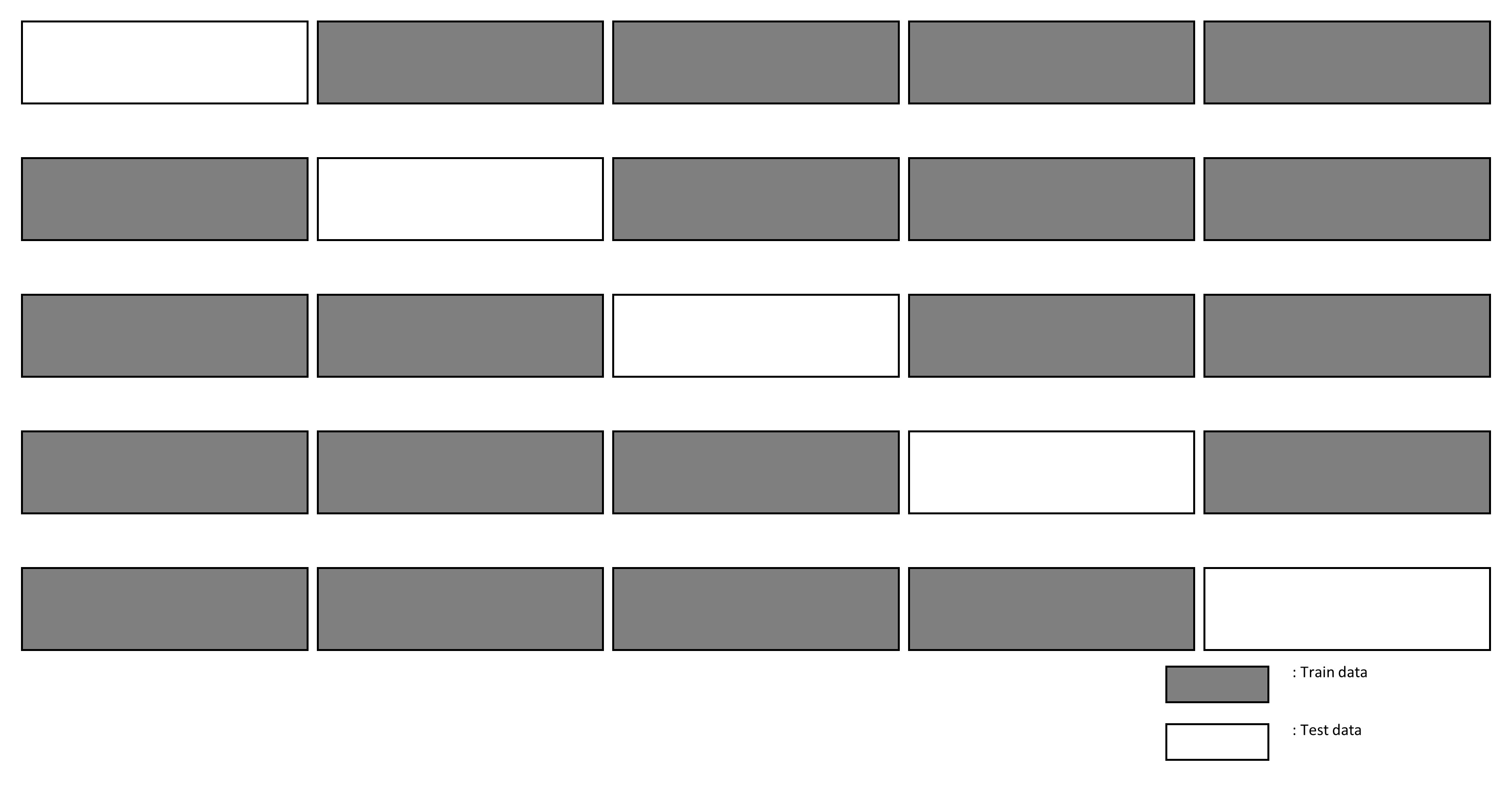}
    \caption{$k$-fold cross validation}
    \label{fig:enter-label}
\end{figure}

\subsection{Model Training and Optimization}
 ~The training dataset was used to optimize the stochastic gradient descent algorithm and the validation dataset was used to validate the performance of model and optimize hyperparameter of the model. The trainable parameters were initialized using the Xavier normal initialization method. To maximize the performance of the model, Bayesian optimization was employed to fine-tune the hyperparameters[22-24]. Specifically, the optimizer and batch size were selected as the key hyperparameters for optimization through this method. By systematically exploring the hyperparameter space, Bayesian optimization aims to identify the optimal configuration that maximizes model performance. The optimization was performed across 10 iterations, varying the optimizer and batch size.

\subsubsection{Training Strategy}
~ To mitigate over-fitting and ensure accurate performance of model, $k$-fold cross validation was employed. As seen in figure 7, $k$-fold cross validation divides the dataset into $N$ subsets, trains on $N-1$ subsets, and uses the remaining 1 subset as the test data. This approach allows using all data for both training and validation, thereby preventing bias in the data and enabling a more accurate evaluation of the model performance. In this study, $k$ was set to 5.

\subsubsection{Bayesian Optimization}
 ~Bayesian optimization is a probabilistic model-based optimization method that designed to find the minimum of an objective function $f$. Firstly, it constructs a surrogate model to approximate the true objective function using exploited data. Then, selects the next point to evaluate based on acquisition function and update the surrogate model. This process continues iteratively to refine the surrogate model and acquisition function to progressively improve optimization outcomes. Acquisition function can be expressed as follows in the equation.

\begin{equation}
    EI(x) = \mathbb{E} [\max (f(x)-f(x^+), 0)]
    \\
    =\left\{\begin{matrix}
 (\mu (x) - f(x^+) - \xi ) \Phi(Z) + \sigma(x) \phi(Z) & \mathrm{if}~\sigma(x) > 0  \\
 0& \mathrm{if} ~ \sigma (x) = 0 \\
\end{matrix}\right.
\end{equation}
\begin{equation}
    Z = \left\{\begin{matrix}
        {{\mu(x)-f(x^+) - \xi} \over { \sigma(x)}} & \mathrm{if}~\sigma(x) > 0 \\
        0 & \mathrm{if}~\sigma(x) = 0\\
    \end{matrix}\right.
\end{equation}

~Bayesian optimization is particularly effective in finding the optimal values especially in a time-effective manner. In this study, Bayesian optimization iterated 10 times.

\section{Results and Discussions}
\subsection{comparative Study}
~To emphasize the advantages of AESLNet, we conducted comparative study from two perspectives. Firstly, analyze the propagation speed-based location prediction method. Secondly, compare the accuracy of AESLNet with several baseline convolutional neural network(CNN) models.

\subsubsection{Propagation Speed-based Location Prediction Method}
~ In section 2.3, we calculated the propagation speed of AE signals and the results are presented as shown in figure 8. The propagation speeds were calculated as follows : an average of 281.59 m/s, a standard deviation of 1058.42 m/s, and a 95\% confidence interval of 61.60 m/s, as shown in Table 1. As shown in figure 8, the propagation speed is not consistent. Particularly, the occurrence of negative propagation speed demonstrates significant non-linearity in wave propagation dynamics within composites. This shows that pinpointing the location of the signal source based on propagation speed is tough.

\begin{figure}[!b]
    \centering
    \includegraphics[width=0.8\linewidth]{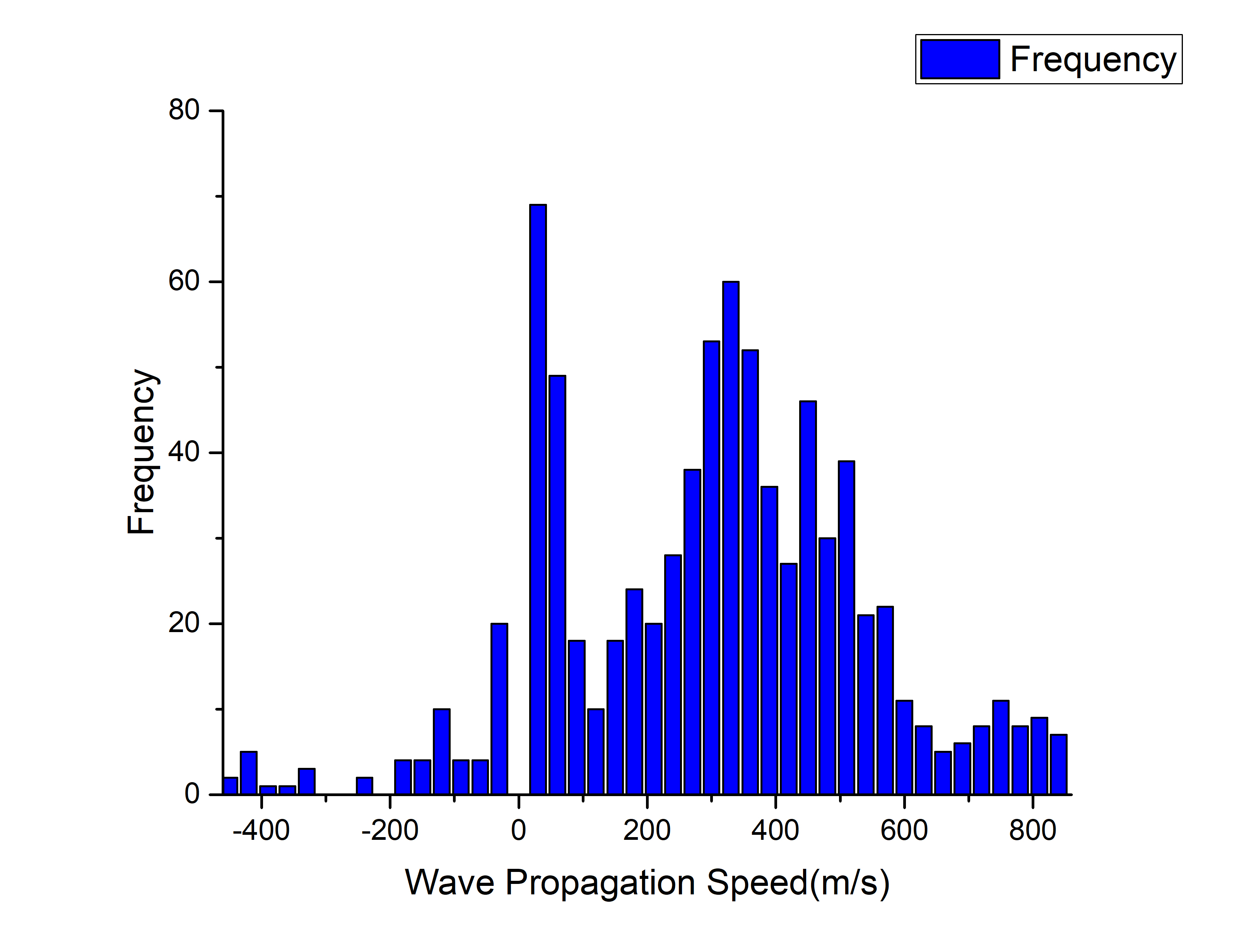}
    \caption{Distribution of calculated wave propagation speed based on TDOA}
    \label{fig:distribution}
\end{figure}

\begin{table}[!b]
\caption{Statistics of calculated wave propagation speed based on TODA(time difference of arrival)}
\label{tab:statistics}
\centering
\begin{tabular}{|c|c|}
\hline
\textbf{Average} &  281.59 m/s  \\ \hline
\textbf{Standard Error} &  33.16 m/s  \\ \hline
\textbf{Standard Deviation} &  1058.42 m/s  \\ \hline
\textbf{Confidence Interval Level(95\%)} & 65.06 m/s   \\ \hline
\end{tabular}
\end{table}

\begin{table}
\caption{Optimization Result}
\label{tab:optimization}
\centering
%\arraystretch{1.5}
\begin{tabular}{|c|c|c|c|}
\hline
\textbf{Model} & \textbf{Optimizer} & \textbf{Batch size} & \textbf{Loss} \\ \hline
VGGNet & SGD & 32 & 0.023 \\ \hline
ResNet & RMSprop & 23 & 0.016 \\ \hline
MobileNet & RMSprop & 23 & 0.025 \\ \hline
AESLNet & RMSprop & 23 & 0.0028 \\
\hline
\end{tabular}
\end{table}

\begin{table}
\caption{Performance of trained AESLNet}
\label{tab:enter-label}
\centering
%\arraystretch{1.5}
\begin{tabular}{|c|c|}
\hline
\textbf{Average Error} & 3.02 mm \\ \hline
\textbf{Standard Deviation} & 2.11 mm \\ \hline
\textbf{Confidence Interval Level(95\%)} & 0.302 mm \\ \hline
\textbf{Maximum Error} & 8.97 mm \\
\hline
\end{tabular}
\end{table}

\subsubsection{Convolutional Neural Network}
~ In this study, 3 CNN models-VGGNet, ResNet, and MobileNet-were used for performance comparison with proposed AESLNet. All models were trained and evaluated using the same methods as described in section 3.2. The result of optimization is shown in table 2. The proposed model, AESLNet outperform the other CNN models. These findings demonstrate that AESLNet, by employing separate convolutional modules for each channels, effectively accommodates the varying signal characteristics arising from difference sensor locations.

\subsection{Quantative Evaluation}
~Performance of trained AESLNet on the train dataset are shown in figure 9(a), with an average error of 3.02 mm, a 95\% confidence interval of 0.302 mm, and a maximum error of 8.97mm, as presented in Table 4. Also the detection results of AESLNet on the validation dataset which are not included in training dataset are shown in figure 9(b). For the validation dataset, an average error of 5.38 mm and a maximum error of 9.89mm were observed.  Given that these errors are less than 10 mm, which is half the grid spacing of 20 mm used in this study, the system achieves a resolution of 20 mm.

\begin{figure}
    \centering
    \begin{subfigure}[b]{0.45\linewidth} % adjust width as needed
        \centering
        \includegraphics[height=0.8\linewidth]{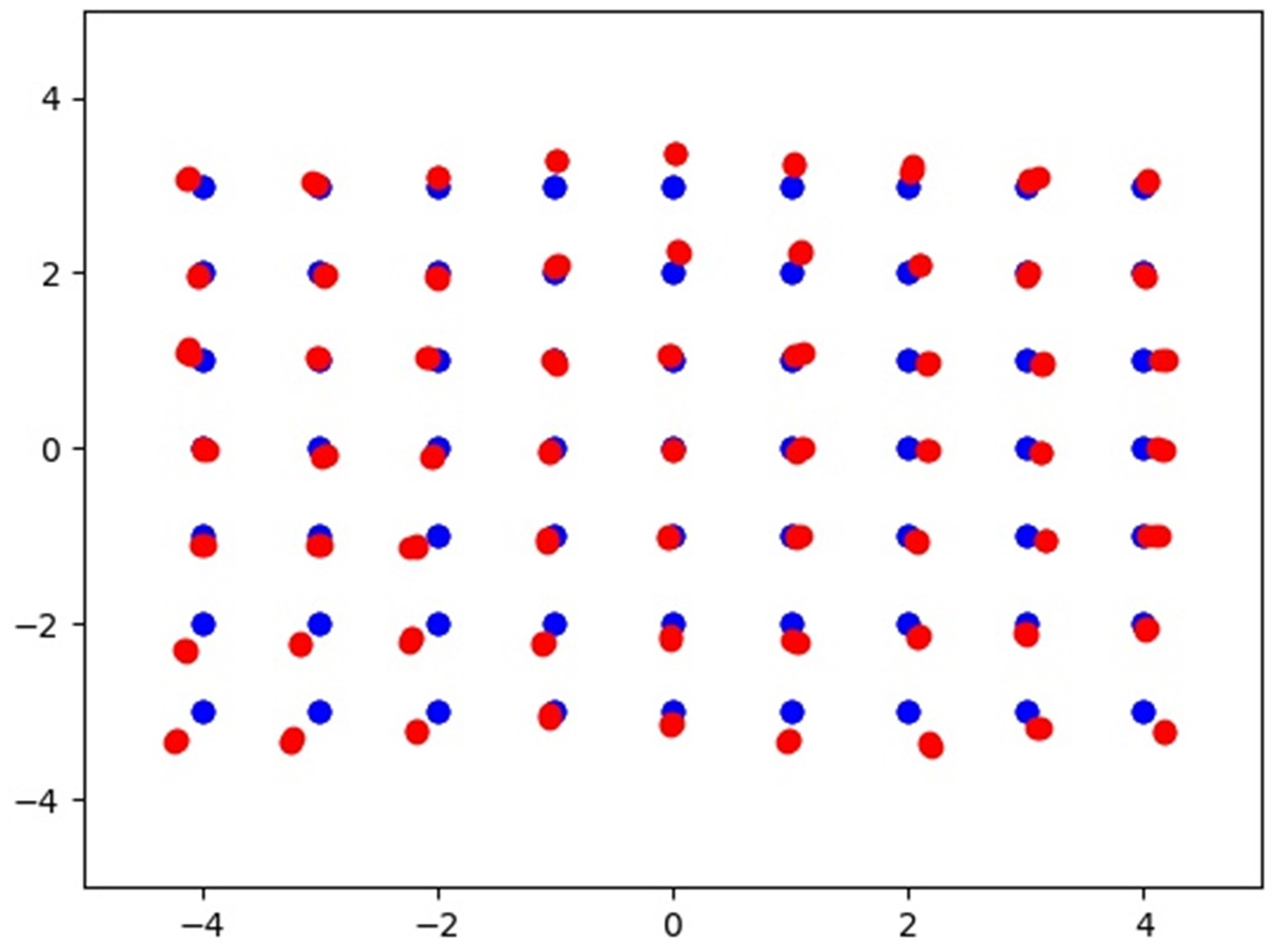}
        \caption{Result for train data}
        \label{fig:train_data}
    \end{subfigure}
    \hfill
    \begin{subfigure}[b]{0.45\linewidth} % adjust width as needed
        \centering
        \includegraphics[height=0.8\linewidth]{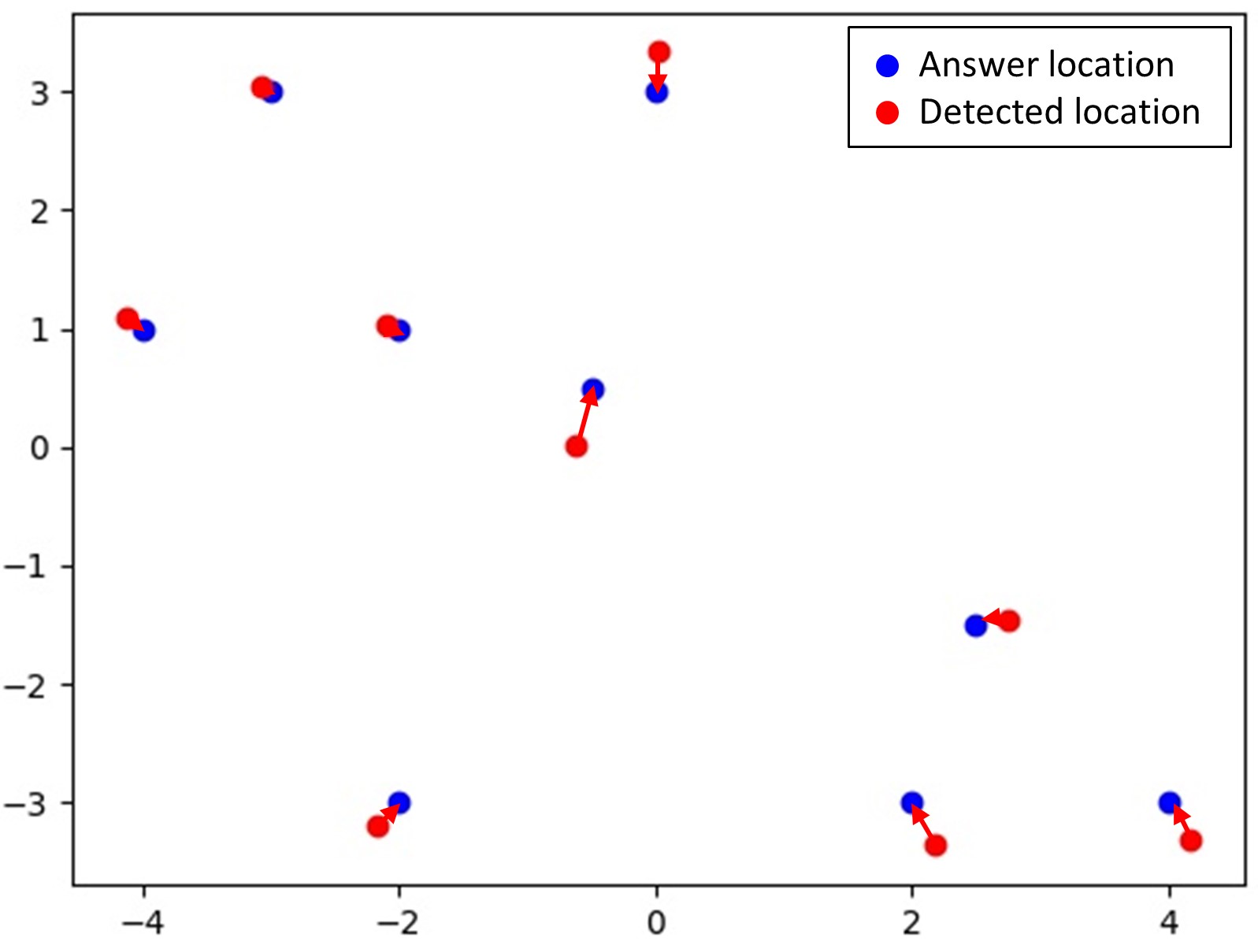}
        \caption{Result for test data}
        \label{fig:test_data}
    \end{subfigure}
    \caption{Detection Results}
    \label{fig:detection_results}
\end{figure}

\section{Conclusion}
~ In this study, we proposed new methodology to localize the source of AE signal in composites called AESLNet.
AE signals were processed through wavelet transform and used as input of AESLNet to localize the acoustic emission sources. To accommodate the differing characteristics of acoustic emission signals arriving at each sensor location, AESLNet was structured with parallel convolutional layers. The detection results were compared with signal arrival time method and three CNN models, leading to the following conclusions.
\\
(1)	To detect the location of AE sources based on time of arrival, we calculated the propagation speed of the AE signals. The results indicated an average speed of 289.59 m/s with a 95\% confidence interval of 65.05 m/s, showing significant variations in propagation speed depending on direction or position. This confirms that localizing methods based on time of arrival are challenging to apply in composites. 
\\
(2)	Comparing the proposed AESLNet with three CNN image recognition mdoels, a discrepancy of approximately 7-fold in loss was observed. This demonstrates that the parallel configuration of convolutional layers in AESLNet effectively captures the varying signal characteristics measured at different sensor positions. Additionally, performance improvements of 93\% were achieved for AESLNet through the optimization of hyperparameters such as the optimizer and batch size.
\\
(3)	In this study, the location of acoustic emission sources was detected using AESLNet, which is designed based on a convolutional neural network. The results showed an average error of 3.02 mm and a maximum error of 8.97 mm. From these findings, it was confirmed that detecting the location of acoustic emission sources in composite materials using convolutional neural networks is feasible with a resolution of 20 mm.
\\
(4)	AESLNet, as proposed in this study, effectively reflects the varying signal characteristics associated with sensor positions, enabling accurate prediction of signal locations in heterogeneous composite materials. Consequently, AESLNet is anticipated to be applicable not only for composite materials but also for detecting signal locations in structures with complex geometries or heterogeneous materials. Furthermore, it is expected that AESLNet can be utilized in damage monitoring technologies for structures that may be exposed to various damage threats.


\begin{thebibliography}{1}

\bibitem{1}
Yoon, D. J., Jeong, J. C., Lee, S. S., and Won, C. W.,
\textit{AE Characteristics of Fatigue Crack Opening and Closure in SWS 490B and Al 7075-T6 alloy},
Trans. Korean Soc. Mech. Eng. A, 2003, Vol. 27, No. 6, pp. 960--968.

\bibitem{2}
Wang, Z. F., Li, J., Ke, W., Zheng, Y. S., Zhu, Z. and Wang, Z. G.,
\textit{Acoustic Emission Monitoring of Fatigue Crack Closure},
Scripta Metallurgica, 1992, Vol. 27, p. 1691.

\bibitem{3}
Mohammadi, R., Najafabadi, M. A., Saghafi, H., Saeedifar, M., and Zarouchas, D.,
\textit{A Quantitative Assessment of the Damage Mechanisms of CFRP Laminates Interleaved by PA66 Electrospun Nanofibers using Acoustic Emission},
Composite Structures, 2021, Vol. 258, No. 113395.

\bibitem{4}
Yang, Y. C., Hwang, W., Park, H. C. and Han, K. S.,
\textit{Vibration Sensing and Impact Location Detection Using Optical Fiber Vibration Sensor},
Key Engineering Materials, 2000, Vol. 187, pp. 661--666.

\bibitem{5}
Kim, E. Y., Kim, M. S., Lee, S. K. and Koh, J. P.,
\textit{CWT-based Method for Identifying the Location of the Impact Source in Buried Pipes},
Trans. Korean Soc. Mech. Eng. A, 2010, Vol. 34, No. 11, pp. 1555--1565.

\bibitem{6}
Salinas, V., Vargas, Y., Ruzzante, J. and Gaete, L.,
\textit{Localization Algorithm for Acoustic Emission},
Physics Procedia, 2010, Vol. 3, pp. 863--871.

\bibitem{7}
Verbis, J. T., Tsinopoulos, S. V., Agnantiaris, J. P. and Polyzos, D.,
\textit{Wave Propagation in Composites},
Recent Advances in Composite Materials, 2003, pp. 35--46.

\bibitem{8}
Zhang, C. Y. and Kim, D. H.,
\textit{Analysis of Impact Position based on Deep Learning CNN Algorithm},
Trans. Korean Soc. Mech. Eng. A, 2020, Vol. 44, No. 6, pp. 405--412.

\bibitem{9}
Tabian, I., Fu, H., Khodaei, Z. and Sharif, Z.,
\textit{Impact Detection on Composite Plates based on convolution neural network},
Key Engineering Materials, 2020, Vol. 827, pp. 476--481.

\bibitem{10}
Ebrahimkhanlou, A. and Salamone, S.,
\textit{Single-Sensor Acoustic Emission Source Localization in Plate-Like Structures Using Deep Learning},
Aerospace, 2018, Vol. 5, Issue. 2.

\bibitem{11}
Sai, Y., Zhao, X., Wang, Lili. and Hou, D.,
\textit{Impact Localizing of CFRP Structure based on FBG Sensor Network},
Photonic Sensors, 2020, Vol. 10, Issue. 1, pp. 88--96.

\bibitem{12}
Hsu, N. N.,
\textit{Acoustic Emissions Simulator},
U. S. Patent US4018084A, Washington D. C., U. S. Patent and Trademark Office, 1976.

\bibitem{13}
Sauge, M. G. R.,
\textit{Investigation of Pencil-Lead Breaks as Acoustic Emission Source},
Journal of Acoustic Emission, 2011, Vol. 29, pp. 184--196.

\bibitem{14}
Al-Jumaili, S. K., Pearson, M. R., Holford, K. M., Eaton, M. J. and Pullin, R.,
\textit{Acoustic emission source location in complex structures using full automatic delta T mapping technique},
Mechanical Systems and Signal Processing, 2016, Vol. 72, pp. 513--524.

\bibitem{15}
Oh, H. T., Won, J. I., Woo, S. C. and Kim, T. W.,
\textit{Determination of Impact Damage in CFRP via PVDF Signal Analysis with Support Vector Machine},
Materials, 2020, Vol. 13, No. 22, pp. 1--23.

\bibitem{16}
Dobrzycki, A., Mikulski, S. and Opydo, W.,
\textit{Using ANN and SVM for the Detection of Acoustic Emission Signals Accompanying Epoxy Resin Electrical Treeing},
Applied Sciences, 2019, Vol. 9, No. 8, pp. 1--14.

\bibitem{17}
Saeedifar, M., Najafabadi, M., Zarouchas, D., Toudeshky, H. H., and Jalalvand, M.,
\textit{Barely Visible Impact Damage Assessment in Laminated Composites using Acoustic Emission},
Composites Part B: Engineering, 2018, Vol. 152, pp. 180--192.

\bibitem{18}
Ince, N., Kao, C. S., Tewfik, A. and Labuz, J. F.,
\textit{A Machine Learning Approach for Locating Acoustic Emission},
Eurasip Journal on Advances in Signal Processing, 2010, Vol. 2010, No. 895486, pp. 1--14.

\bibitem{19}
Simonyan, K. and Zisserman, A.,
\textit{Very Deep Convolutional Networks for Large-scale Image Recognition},
arXiv 1409.1556, 2015.

\bibitem{20}
He, K., Zhang, X., Shaoqing, R. and Sun, J.,
\textit{Identity Mappings in Deep Residual Networks},
Computer Vision – ECCV2016, 2016.

\bibitem{21}
Howrad, A. G., Zhu, M., Chen, B., Kalenichenko, D., Wang, W., Weyand, T., Andreetto, M. and Adam, H.,
\textit{MobileNets: Efficient Convolutional Neural Networks for Mobile Vision Applications},
ArXiv, 2017.

\bibitem{22}
Feurer, M. and Hutter, F.,
\textit{Hyperparameter Optimization},
Automated Machine Learning, Springer, Cham, 2019, pp. 3--33.

\bibitem{23}
Tran, N., Schneider, J., Weber, I. and Qin, A. K.,
\textit{Hyper-parameter Optimization in Classification: To-do or not-to-do},
Pattern Recognition, Vol. 103, 2020, No. 107245.

\bibitem{24}
Snoek, J., Larochelle, H. and Adams, R. P.,
\textit{Practical Bayesian Optimization of Machine Learning Algorithms},
Advances in Neural Information Processing Systems, Vol. 4, 2012, pp. 2951--2959.


\end{thebibliography}
\end{document}